
\magnification=1200
\hsize 15.0 true cm
\vsize 23.0 true cm
\def\c{\centerline}
\def\v{\vskip 1pc}
\def\ej{\vfill\eject}
\def\r{\vec r}
\def\ra{\rangle}
\def\re{{\rm e}}
\def\s{\sigma}
\def\G{\Gamma_0}
\def\tr{{\rm tr}}
\def\half{{1 \over 2}}
\def\da{\dagger}
\overfullrule 0 pc
\
\vskip 1.5pc
\parindent 3pc\parskip 1pc
\c{\bf  Changes in the radius of a nucleon in interaction with another
nucleon}
\vskip 4 pc
\c{G. K\"albermann}
\v
\c{Rothberg School for Overseas Students and Racah Institute of Physics}
\c{Hebrew University, 91904 Jerusalem, Israel}
\v
\c{and}
\v
\c{L.L. Frankfurt and J.M. Eisenberg}
\v
\c{School of Physics and Astronomy}
\c{Raymond and Beverly Sackler Faculty of Exact Sciences}
\c{Tel Aviv University, 69978 Tel Aviv, Israel}
\vskip 2pc
{\bf Abstract:-}  We consider a two-nucleon system described by 
two different skyrmion models that provide attraction for the
central $NN$ potential.  One of these models is based on the product 
ansatz and the other on dilaton coupling.  Within these models we 
ask the question, To what degree does the nucleon swell or shrink 
when the internucleon separation distance is appropriate to
attraction or repulsion?  We find typically swelling of 3 to 4
percent for central attraction of some 40 to 50 MeV.

\vfill

\noindent February, 1994.

\ej

\baselineskip 15 pt
\parskip 1 pc \parindent 3pc

A major goal of nuclear physics in the years since the establishment
of quantum chromodynamics (QCD) as the preferred theory of strong 
interactions has been to study the mutual 
influences between nucleon substructure and the behavior of complex
nuclei.  Well over ten years ago, it was suggested [1] that
there might be appreciable changes of the nucleon radius when the 
nucleon is in a nuclear medium.  This issue was then explored in a
number of papers [2-7]; the situation as of 1988 is reviewed in ref. 
[8]. Early arguments [2-5] were in part based on the expectation that
$r\,M = r_0\,M_0,$ where $r$ and $M$ are the nucleon radius and
mass within the nuclear medium, and $r_0$ and $M_0$ are these same
two quantities for a free nucleon.  However, there are many
possible sources for the ``effective'' features in the 
effective mass $M,$ ranging from nonrelativistic many-body physics
to issues of relativistic mean fields, while the changes of the 
nucleon radius in the hadronic medium are expected to arise mainly
from the polarization of the particle.  
Put in other terms, the issue of the nucleon effective mass in the
medium is likely to relate most strongly to its behavior as a 
quasiparticle, that is to say, to the dressing that is supplied by 
its interaction with the other nucleons around it, independent of
questions of nucleon substructure; the effective mass, for example,
is quite naturally a long-range phenomenon in the nucleus.
On the other hand, we tend to view changes in nucleon radius as mainly a 
result of responses of the internal nucleon constituents to their
immediate
surroundings, and thus as a short-range effect.  Further, the empirical
limits on the changes of the nucleon radius in the nucleus suggest 
[8] that $|r/r_0 -1| \le 0.04,$ roughly, while the nucleon effective
mass may differ by a few times this amount according to many
estimates of $M/M_0,$ so that it is difficult to accept the
simple view that $r \sim 1/M.$

Since the understanding of the $NN$ force obtained from models intended
to reflect QCD physics also suggests that the central attraction
in that force derives from polarization phenomena, it has become
increasingly interesting to attempt to link those two polarization
features directly [6-8].  The approach to the $NN$ problem based on 
skyrmions lends 
itself particularly readily to such studies.  (The use of skyrmions
for nuclear problems has by now been surveyed by nearly all the
practitioners in the field; we note here the reviews [9-13] that
are particularly close to the issues raised in this paper.)
There one finds---much as in the application of the nonrelativistic 
quark model to the $NN$ force problem [14]---that it is crucial to
include the polarization of each individual nucleon in order to 
get an attractive central potential of medium range.  This is
perhaps best handled by allowing for the distortion of the full 
two-skyrmion solution as the two baryons approach each other [12,13],
but very similar effects may also be dealt with by the explicit
admixture of nucleon excitations such as the $\Delta(1232)$ and
$N(1440)$ as the internucleon separation distance $R$ is decreased 
[11].  Some skyrmion studies have indeed been made of the 
changes in nucleon radius as $R$ is varied [15,16], but these were
carried out while the search for a fuller understanding of possible
sources of attraction in the skyrmion approach was at its peak. As a 
result, the models in question in fact contained no attraction, 
and therefore merely
confirmed that when skyrmions are mainly engaged in repelling each
other they also cause each other to shrink.\footnote{*}{There also 
exists an interesting skyrmion study [17] of radius changes in an 
infinite hadronic medium.}  As was early pointed out [6-8],
there are very general arguments that imply that an attractive
potential will be accompanied by nucleon
swelling, and it is the purpose of this paper to study
this at a quantitative level using skyrmions.

We first sketch some of the relevant formalism,
beginning with the Skyrme model using the product ansatz.
The skyrmion is here described by the lagrangian 
$$\eqalign{{\cal L} & = -{F_\pi^2 \over 16} \tr(L_\mu  L^\mu )
+ {1 \over 32 e^2} \tr[L_\mu,L_\nu]^2 \cr
& + {\gamma \over 8 e^2} \tr (L_\mu L_\nu)^2
- {\epsilon^2 \over 8} (\tr B^\mu)^2,}\eqno(1)$$
with
$$L_\mu \equiv U^\da\partial_\mu U\eqno(2)$$
and
$$B^\mu \equiv -{\epsilon^{\mu\alpha\beta\gamma} \over 24\pi^2}
\tr(L_\alpha L_\beta L_\gamma),\eqno(3)$$
where $U(\r,t)$ is the unitary SU(2) chiral field, $F_\pi$ is the
pion decay constant (with experimental value 186 MeV), $e$ is
the Skyrme parameter, and $\gamma$ and $\epsilon$ are coefficients
of additional attractive and repulsive terms, respectively [11].
(We note that these two terms are necessary in order to obtain
medium-range attraction in the $NN$ potential in the present
approach.)
The static solution for the $B = 1$ case is the well-known hedgehog 
$$U(\r) = \exp[i\vec\tau\cdot\hat{\r}\, F(r)], \eqno(4)$$
where $F(r)$ is the profile function or chiral angle.  For the $B
= 2$ soliton, we use the product ansatz [11]
$$U_{B=2} = A(t)U(\r-\r_1)A^\dagger(t)\, B(t)U(\r-\r_2)B^\dagger(t),
\eqno(5)$$
where $\r_1$ and $\r_2$ are the baryon locations, and $A(t)$ and 
$B(t)$ are time-dependent rotations in SU(2).  Introducing
eq. (5) into the hamiltonian derived from eq. (1) and subtracting
the one-body energies, we find the two-body potential.

Attraction for the $NN$ system is achieved within the product ansatz 
(5) by carrying out a further variational calculation [11] in which
the baryon resonances $\Delta(1232)$ and $N^*=N(1440)$ are admixed by
$$\eqalign{|\widetilde{NN}(R)\ra & = \alpha(R)|NN(R)\ra
+\beta(R)|N\Delta(R)\ra +\gamma(R)|\Delta N(R)\ra \cr
& +\epsilon(R)|\Delta\Delta(R)\ra
+\zeta(R)|NN^*(R)\ra+\eta(R)|N^*N(R)\ra \cr
& +\theta(R)|N^*N^*(R)\ra\cdots.}
\eqno(6)$$
(The ellipsis refers to the possible inclusion of, say, the 
$\Delta(1670),$ whose contribution proves to be negligible here.)
That is to say, at each internucleon separation distance $R$ we
minimize the energy for the two-nucleon system with respect to the
coefficients $\alpha(R)$ through $\theta(R),$ etc., after we have 
obtained this energy from projected states for the nucleon and the
admixed baryon resonances.  The Roper $N(1440)$ is
handled as a vibrating breathing
mode of the nucleon (see [11] and references therein for details). 
Changes in the radius of the nucleon while in interaction with a
second nucleon in this description are studied by taking the second
nucleon as a spectator and evaluating the root-mean-square radius
for the first.

We now turn to the Skyrme model involving coupling to a dilaton.
It was early realized [18] that such an approach allows
for the incorporation of the QCD trace anomaly.
Subsequently such models were studied for their possible advantages
in obtaining medium-range central attraction [19,20].  The
introduction of a new length scale through the dilaton has the
effect of inducing a sharper edge for the skyrmion, which now
exists with bag-like support within the dilaton.  This seems an 
appealing and natural way to cut off the long tail of skyrmion
repulsion, generated by the same mechanism that stabilizes the 
skyrmion, and thus to enhance attraction at medium ranges.  We
have thus found this model to be interesting in terms of its
predictions for nucleon radius change as well.

Towards such a study we take the usual [18-20] lagrangian for a
skyrmion coupled with a dilaton,
$$\eqalign{{\cal L}  = {\cal L}_{\rm sym} - V(\s) 
& = \re^{2\s}[\half\G^2\partial_\mu\s\partial^\mu\s
- {F_\pi^2 \over 16}\tr(L_\mu L^\mu)] \cr
& +{1 \over 32 e^2}\tr[L_\mu,\, L_\nu]^2 - V(\s).}\eqno(7)$$
Here ${\cal L}_{\rm sym}$ preserves scale invariance and $V(\s)$ is 
determined [18] from the trace anomaly to be
$$V(\s)={C_G \over 4}[1 + \re^{4\s}(4\s-1)],\eqno(8)$$
and there are two new constants $\G$ and $C_G.$
The product ansatz of eq. (5) is now augmented by an assumption of
additivity for the dilaton $\s$-field,
$$\s_{B=2}=\s_1+\s_2.\eqno(9)$$
Once again the $NN$ potential is identified by subtracting the energy for
the $B=2$ system with separation $R \to \infty$ from that with finite 
separation $R.$ 

The nucleon radius modification is this time evaluated by minimizing 
the total static energy (without rotational energy),
$$E(X) = {1 \over \lambda} E_{-1}(X) + \lambda E_1(X) 
+ {1 \over \lambda^3} E_{-3}(X),\eqno(10)$$
where $X = \lambda R,$  with 
respect to a change in the length scale $\lambda.$ 
In eq. (10) we identify the various contributions to the static
energy according to their behavior under $\r \to \lambda \r,$ namely,
$$E_{-1}(X) = \int \re^{2\s}\bigg[{\Gamma_0^2 \over 2}(\nabla\s)^2
- {F_\pi^2 \over 16} \tr (L_i L_i)\bigg]{\rm d}{\vec x},\eqno(11a)$$
$$E_1(X) = \int {1 \over 32e^2} \tr [L_i,L_j]^2 {\rm d}{\vec x},\eqno(11b)$$
and
$$E_{-3}(X) = \int V(\s) {\rm d}{\vec x},\eqno(11c)$$
where $\vec x=\lambda \r.$  The resulting minimum then fixes the
preferred scale for the interacting system, and with it a slightly modified
two-nucleon potential.  That is to say, the scale invariance of 
${\cal L_{\rm sym}}$ in the original dilaton-skyrmion lagrangian of eq. (7)
has been broken by the additivity ansatz for the dilaton, eq. (9), and
our restricted minimization with respect to $\lambda$ then provides an
approximate solution to the full $B = 2$ dilaton-skyrmion problem which 
improves upon the simple product-plus-addditivity assumptions of eqs. (5) 
and (9).

For the usual Skyrme model with hedgehog and product ansaetze of eqs. (1)
through (5) we show in fig.~1 the central $NN$ potential and ratio of the 
interacting-nucleon radius to the free-nucleon radius.  The central
potential in this case reaches an attraction of about -11 MeV and the
corresponding change in the nucleon radius in interaction is about 
4 percent.  The dilaton case is shown in fig.~2, and yields
considerably more attraction, reaching possibly overly-large values deeper
than -40 MeV.  Nonetheless, the radius changes are again a rather modest
3 percent or so.  In both models, as $R$ becomes very small the two 
nucleons have large overlap, strong repulsion sets in, the nucleons
begin to show shrinking, and the entire use of the product ansatz
quickly becomes meaningless.  In the case of the dilaton, the additivity
assumption, eq. (9), is also bad for small $R.$  For large $R,$ the
skyrmion shows an unrealistically long tail of interaction between the 
two nucleons, reflected both in the potential $V$ and in the deviation of
$r/r_0$ from unity.  (Since both of these quantities have been calculated
on a rather sparse grid---seen in the nexuses of the straight-line
segments---there is a lack of smoothness in the results shown
here.)

Both cases show that, as the potential moves from large positive
values at small $R$ through zero and on to attraction, the nucleon
changes from a shrunken conditon to a swollen one (though in neither 
situation do the changeovers of $V$ and of $r/r_0$ occur at precisely 
the same separations).  There is thus a clear link in two rather 
different skyrmion 
descriptions of the two-nucleon system between attraction and nucleon
swelling.  In both cases the nucleon swelling is quit modest, and well
within the limits set by present experiment [8].  Shrinking sets in
for $R <$ 1.5 fm or so, and it is in this region that quark-gluon
degrees of freedom may first be required in the description of hadronic 
systems (e.g., neutron stars).  Since we have
calculated only for the $B=2$ system, there is here little or no
question of a link between nucleon swelling or shrinking and an 
effective mass within a hadronic medium as this is normally understood.
To the degree that the skyrmion is a valid description for the range
1 fm $\le R \le$ 3 fm, our results span the region from large separation, 
where the effects involve overall hadronic behavior, down to small $R,$ 
where perturbative QCD enters.  The changes in nucleon radius for small 
internucleon separation should be enhanced in measurements of nuclear
form factors since these will be sensitive to the third power of the
scaling factor for moderate momentum transfer.  We note that, since we
expect, on the whole, nucleon shrinking for small $R$ and swelling for
intermediate $R,$ the effects in form factors are likely to show
changes as different ranges of momentum of the nucleon within the
Fermi sea are probed.  (For heavy nuclei, the swelling at intermediate
ranges of $R$ may receive partial compensation from the suppression
of the pion cloud around the bound nucleon [21].)
One may hope that with the arrival of new data on high-energy electron 
scattering, for example from CEBAF, the study of changes in the nucleon
radius while the particle is in interaction may serve as an additional 
tool to unravel the delicate question of mutual nucleon polarization
at medium-range separations.

This research was supported in part (G.K. and J.M.E.) by the Israel 
Science Foundation, in part (L.L.F.) by the U.S.-Israel Binational Science 
Foundation, and in part (J.M.E.) by the Yuval Ne'eman 
Chair in Theoretical Nuclear Physics at Tel Aviv University.
Parts of it were carried out while J.M.E. was a guest
at the Institute for Theoretical Physics of the University of Frankfurt,
and he would like to thank Professor Walter Greiner there for his very 
kind hospitality.

\vskip 2 cm

\noindent {\bf References:-}
\v
\baselineskip 12pt
\parskip 0pc
\parindent 1pc
\hangindent 2pc
\hangafter 10

\item{1.}  J.V. Noble, Phys. Rev. Lett. {\bf 46} (1981) 412 
and Phys. Lett. B {\bf 178} (1986) 285.
\v
\item{2.}  L.S. Celenza, A. Rosenthal, and C.M. Shakin, Phys.
Rev. Lett. {\bf 53} (1984) 892 and Phys. Rev. C {\bf 31} (1985) 
232.
\v
\item{3.}  P.J. Mulders, Phys. Rev. Lett. {\bf 54} (1985) 2560
and Nucl. Phys.  {\bf A459} (1986) 525.
\v
\item{4.}  T.D. Cohen, J.W. Van Oerden, and A. Picklesimer,
Phys. Rev. Lett. {\bf 59} (1987) 1267.
\v
\item{5.}  M.K. Banerjee, Phys. Rev. C {\bf 45} (1992) 1359.
\v
\item{6.}  L. Frankfurt and M. Strikman, Nucl. Phys. B {\bf 250}
(1985) 123.
\v
\item{7.}  M. Oka and R.D. Amado, Phys. Rev. C {\bf 35} (1987) 1586.
\v
\item{8.}  L. Frankfurt and M. Strikman, Phys. Repts. {\bf 160}
(1988) 235.
\v
\item{9.} I. Zahed and G.E. Brown, Phys. Repts. {\bf 142} (1986) 1.
\v
\item{10.} G. Holzwarth and B. Schwesinger, Repts. Prog. Phys. {\bf 49}
(1986) 825.
\v
\item{11.} J.M. Eisenberg and G. K\"albermann, Prog. Part. Nucl. Phys. 
{\bf 22} (1989) 1.
\v
\item{12.} T.S. Walhout and J. Wambach, Int. J. Mod. Phys. E {\bf 1}
(1992) 665.
\v
\item{13.} M. Oka and A. Hosaka, Ann. Rev. Nucl. Part. Sci. {\bf 42}
(1992) 333.
\v
\item{14.}  K. Maltman and N. Isgur, Phys. Rev. D {\bf 29} (1984) 952.
\v
\item{15.}  M. Oka, K.F. Liu, and H. Yu, Phys. Rev. D {\bf 34} 
(1986) 1575.
\v
\item{16.}  G. K\"albermann and J.M. Eisenberg, Phys. Lett. B 
{\bf 188} (1987) 311.
\v
\item{17.}  I.N. Mishustin, Sov. Phys. JETP {\bf 71} (1990) 21
[Zh. Eksp. Teor. Fiz. {\bf 98} (1990) 41].
\v
\item{18.}  J. Schechter, Phys. Rev. D {\bf 21} (1980) 3393;

H. Gomm, P. Jain, R. Johnson, and J. Schechter, Phys. Rev. D {\bf 33}
(1986) 801.
\v
\item{19.}  H. Yabu, B. Schwesinger, and G. Holzwarth, Phys. Lett. B
{\bf 224} (1989) 25.
\v
\item{20.}  K. Tsushima and D.O. Riska, Nucl. Phys. {\bf A560} (1993)
985.
\v
\item{21.}  B.L. Friman, V.R. Pandharipande, and R.B. Wiringa,
Phys. Rev. Lett. {\bf 51} (1983) 763.
\ej

\parindent 0 pc
{\bf Figure captions:-}
\v\v
Fig. 1.  The $NN$ potential and ratio of the radius for the 
interacting nucleon to that of the free nucleon $r/r_0$
for the usual skyrmion with product ansatz and baryon-resonance
admixtures, eqs. (1) through (6).
The parameters in eq. (1) are taken to have the values $F_\pi=130$ MeV,
$e=20,$ $\gamma=0.50,$ $\epsilon=2.58,$
and $m_\pi=139$ MeV,  known from earlier studies [11] to
yield a fair amount of attraction in the central potential.  They
produce the masses $M_N=998$ MeV, $M_\Delta=
1211$ MeV, and $M_{N^*}=1270$ MeV for the nucleon,
$\Delta,$ and Roper.  
\v
Fig. 2.  The $NN$ potential and ratio of the radius for the 
interacting nucleon to that of the free nucleon $r/r_0$
for the skyrmion coupled to a dilaton, eqs. (7) through (11).
The parameters used were $\Gamma_0 = 306$ MeV, $C_G = 
(121\ {\rm MeV})^4,$ and $F_\pi = 186$ MeV, which produce
$M_N = 1335$ MeV.  No attempt was made here to search for
more realistic parameters since our interest was only in the link
between attraction/repulsion and swelling/shrinking.
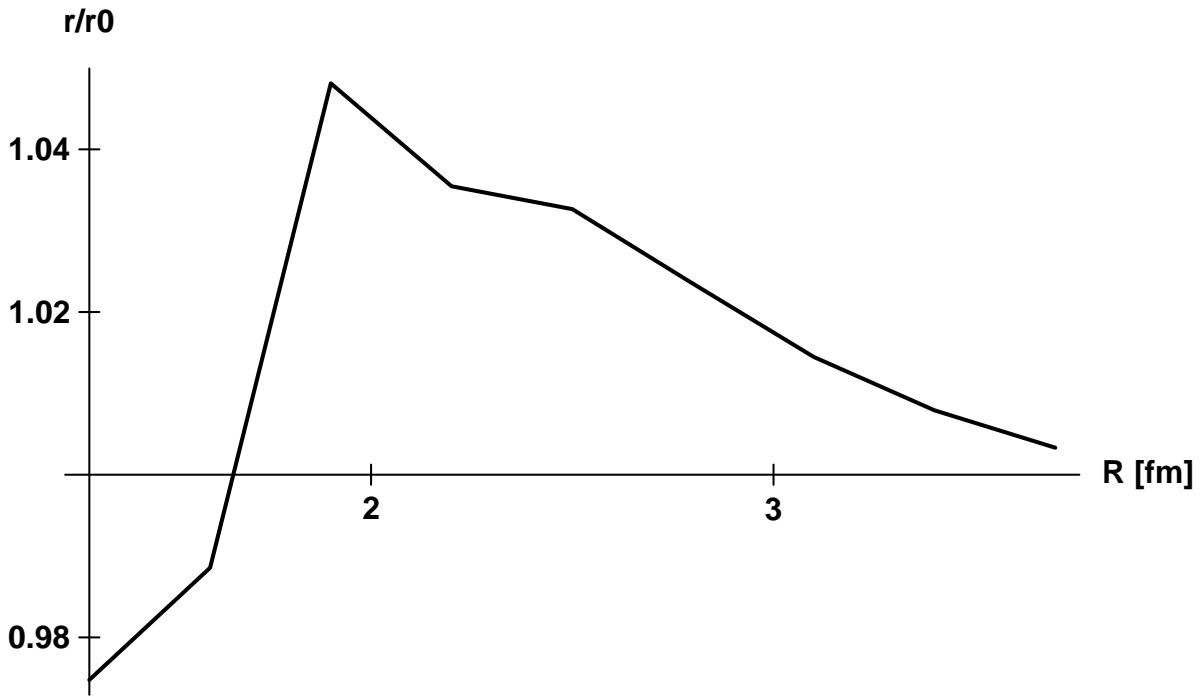
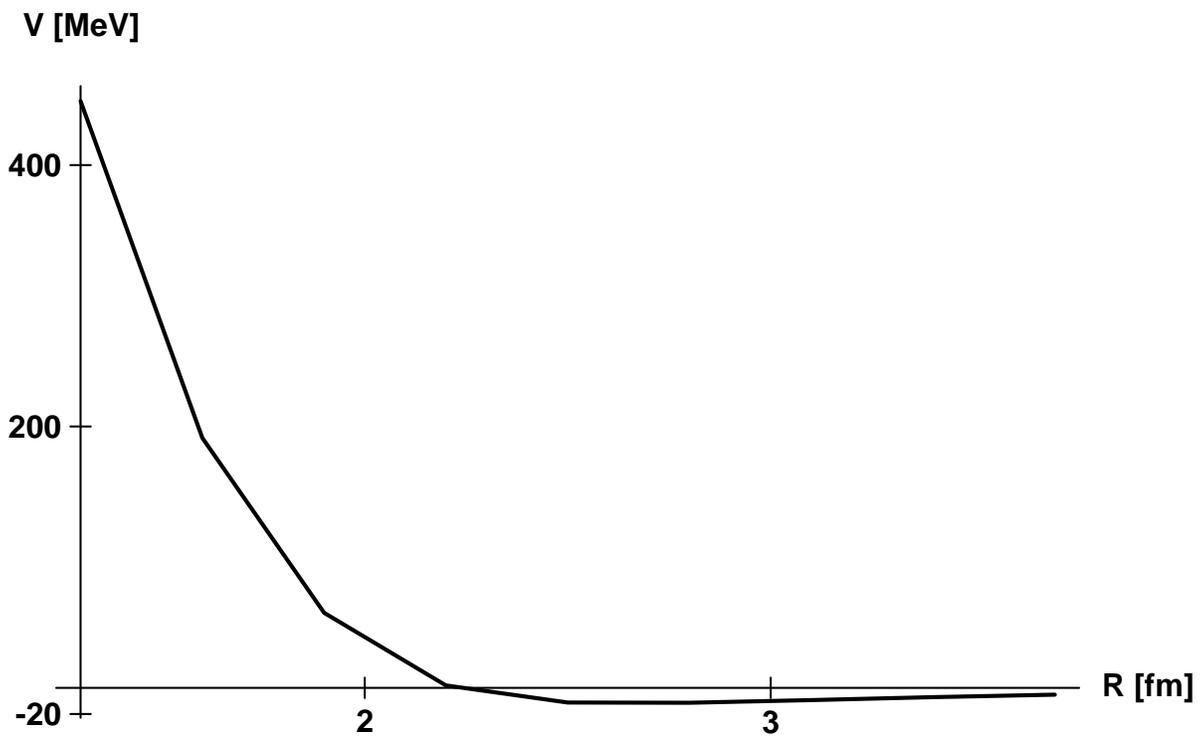
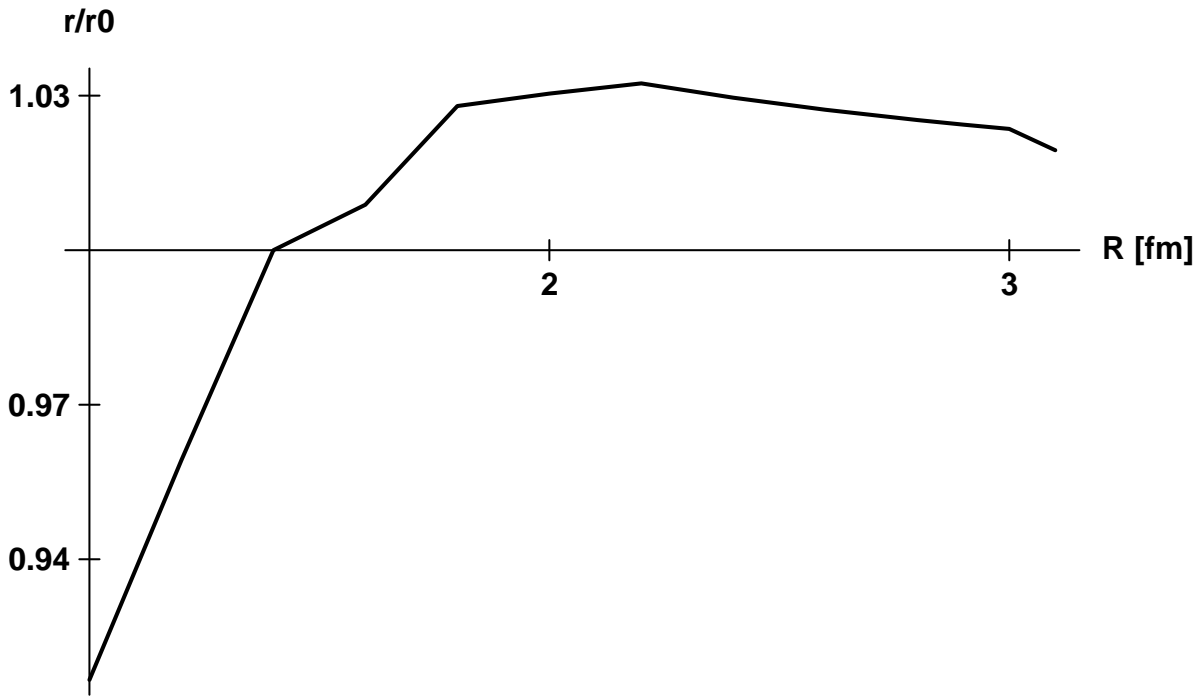
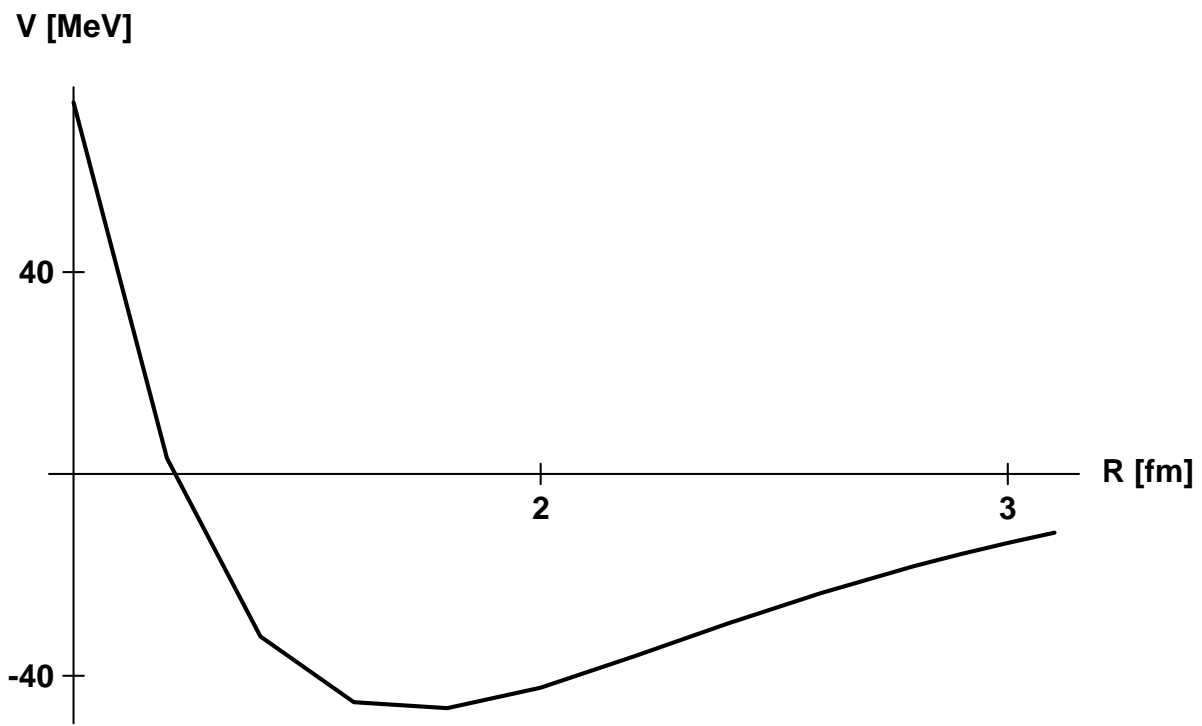
\bye